\documentstyle[prc,aps,preprint,tighten]{revtex}
\begin{document}
\preprint{                   IASSNS-AST 95/47}
\draft
\title{          Standard Neutrino Spectrum from $^8$B Decay}
\author{               John N.~Bahcall and E.~Lisi%
\thanks{Also Dipartimento di Fisica and Sezione INFN di Bari, Bari, Italy.}}
\address{  Institute for Advanced Study, Princeton, New Jersey 08540}
\author{                      D.~E.~Alburger}
\address{  Brookhaven National Laboratory, Upton, New York 11973}
\author{                     L.~De Braeckeleer}
\address{Department of Physics, University of Washington, Seattle,
                              Washington 98195}
\author{                      S.~J.~Freedman}
\address{Department of Physics and Lawrence Berkeley National Laboratory, \\
            University of California, Berkeley, California 94720}
\author{                       J.~Napolitano}
\address{   Rensselaer Polytechnic Institute, Troy, New York 12180} 
\maketitle
\begin{abstract}
We present a systematic evaluation of the shape of the neutrino energy 
spectrum produced by beta-decay of $^8$B. We place special emphasis on 
determining the range of uncertainties permitted by existing laboratory data 
and theoretical ingredients (such as forbidden and radiative corrections). 
We review and compare the available experimental data on the 
$^8$B$(\beta^+){}^8$Be$(2\alpha)$ decay chain. We analyze the theoretical and 
experimental uncertainties quantitatively. We give a numerical representation 
of the best-fit (standard-model) neutrino spectrum, as well as two extreme 
deviations from the standard spectrum that represent the total (experimental 
and theoretical) effective $\pm3\sigma$ deviations. Solar neutrino experiments
that are currently being developed will be able to measure the shape of the 
$^8$B neutrino spectrum above about 5 MeV. An observed distortion of the $^8$B
solar neutrino spectrum outside the range given in the present work could be 
considered as evidence, at an effective significance level greater than three 
standard deviations, for physics beyond the standard electroweak model. We use
the most recent available experimental data on the Gamow--Teller strengths in 
the $A=37$ system to calculate the $^8$B neutrino absorption cross section on
chlorine:  $\sigma_{\rm Cl}=(1.14\pm0.11)\times10^{-42}$~cm$^2$ ($\pm3\sigma$ 
errors). The chlorine cross section is also given as a function of the 
neutrino energy. The $^8$B neutrino absorption cross section in gallium is 
$\sigma_{\rm Ga}=(2.46^{+2.1}_{-1.1})\times10^{-42}$ cm$^2$ 
($\pm3\sigma$ errors).\\
\end{abstract}
\pacs{PACS number(s): 96.60.Kx, 23.40.Bw, 23.60.+e, 27.20.+n}
%
%
\section{Introduction}

The solar neutrino spectrum is currently being explored by four 
underground experiments: the pioneering Homestake (chlorine)
detector \cite{Da94}, the Kamiokande (water-Cherenkov) detector
\cite{Su94}, and two gallium detectors, GALLEX \cite{An95} 
and SAGE \cite{Ni94}.

These first-generation experiments have  shown that the observed 
solar neutrino event rates are  lower than expected \cite{Ba95}, 
giving rise to ``solar neutrino problems'' \cite{Ba94} that cannot be 
solved within the standard experimental and theoretical understanding 
of the physics of the sun and of the electroweak interactions.

Bahcall \cite{Ba91} has shown that the $^8$B$(\beta^+){}^8$Be (allowed)
decay produces the same shape for the ($^8$B) neutrino spectrum,
up to gravitational redshift corrections of ${\cal O}(10^{-5})$,
independent of whether the neutrinos are created in a terrestrial 
laboratory or in the center of the sun. Thus,  experimental evidence 
for a deviation of the $^8$B solar neutrino spectrum from the laboratory
shape would  constitute evidence for physics beyond the standard model 
of the electroweak interactions. Indeed, powerful new experiments, such 
as the SuperKamiokande \cite{SKam}, the SNO \cite{Sudb}, and the ICARUS 
\cite{ICAR} detectors, will measure the {\em shape\/} of the high-energy 
($E_\nu\gtrsim5$ MeV) part of the solar neutrino spectrum, which 
originates from the beta-decay of $^8$B produced in the sun.

The calculation of the $^8$B solar neutrino spectrum dates back to 1964, 
when it was pointed out  \cite{Ba64} that the usual $\beta$-decay 
allowed spectrum  should be averaged over the intermediate $2^+$ states
of $^8$Be, as derived by the subsequent alpha-decay $^8$Be$(2\alpha)$. 
Experimental evidence for this smearing was  found  23 years later in 
the associated positron spectrum \cite{Na87}. The calculation of the 
spectrum has been continually improved by Bahcall and collaborators
in \cite{Ba64,Ba78,Ba86}, to which we refer the conscientious reader
for all the details not reported here. The independent calculation of 
the $^8$B neutrino spectrum by J.~Napolitano {\em et al.\/} \cite{Na87}
also compares well with the results in \cite{Ba86}.

In this work,  the evaluation of the $^8$B solar neutrino spectrum is 
further improved, using recently available experimental data and new 
theoretical calculations. Moreover, the maximum  uncertainties 
($\pm3$ effective standard deviations) that can affect the best estimated 
spectrum are determined. The inferred spectral shape is presented 
numerically and graphically, in forms useful for fits to experimental 
results and for phenomenological analyses. The neutrino spectra, 
together with new data  on the Gamow--Teller strengths in the $A=37$ system,
are  used to improve the calculations of the $^8$B neutrino absorption
cross sections for chlorine and for gallium.

The paper is organized as follows. In Sec.~II, the available experimental 
data on the  $^8$B$(\beta^+){}^8$Be$(2\alpha)$ decay chain are reviewed 
and discussed, in order to extract an optimal data set and in order to 
evaluate the experimental uncertainties. In Sec.~III the best neutrino 
spectrum is calculated, including the appropriate  radiative corrections.
The total (theoretical and experimental) uncertainties are used to 
calculate a ``$+3\sigma$'' and a ``$-3\sigma$'' spectrum, characterizing
the maximum deviations from the optimal spectrum. In Sec.~IV these 
spectra are applied to a refined calculation of the $^8$B neutrino
absorption cross sections for $^{37}$Cl and for $^{71}$Ga. A summary of 
the work is presented in Sec.~V.

\section{Experimental data on the
         $^8$B$(\beta^+){}^8$B\lowercase{e}$(2\alpha)$ decay chain}

In this section, we  present  a compilation of the available experimental 
data on beta and alpha decay of $^8$B. Then the various $\alpha$-decay 
data are compared and used to fit the $\beta$-decay data. The best-fit 
alpha spectrum and its $\pm3\sigma$ range emerge naturally from this 
overconstrained comparison.

The $\beta$-decay of $^8$B populates $2^+$ states in $^8$Be, which then 
breaks up into two $\alpha$-particles \cite{Aj88}. The energy levels are 
shown  in Fig.~1. The  $\beta$, $\alpha$ and correlated $\beta$-$\alpha$ 
spectra have been investigated by several experimental groups. In 
addition to the solar neutrino application, this decay chain is of 
special interest because  the $A=8$ nuclear isotriplet (Li,~Be,~B) can 
be used to search for violations of the conserved vector current (CVC) 
hypothesis or for the existence of second-class currents (SCC) 
(see, e.g., the review \cite{Gr85}), and is also an interesting subject 
for  $R$-matrix analyses \cite{Ba69,Wa86,Ba89}. 

For the determination of the $^8$B (solar) neutrino spectrum, a single 
precise determination of the alpha spectrum is, in principle, all that 
is required. In practice, we take advantage of the redundancy of the 
available experimental information to estimate both a  preferred alpha
spectrum  and the possible uncertainties.

\subsection{Beta decay data}

The most recent determination of $^8$B $\beta$-decay spectrum is 
reported (graphically) in \cite{Na87}. The original  number of detected 
counts in each of the 33 channels used (for 
momentum $p \gtrsim 9$ MeV) have been made available by one of us (J.N.).
We discuss below in  detail the energy calibration of the $\beta$ decay
data of \cite{Na87}, for its particular relevance to the present work.

The calibration procedure used in \cite{Na87} was similar to that in
\cite{Ca90}, which employed the same spectrograph to determine the weak 
magnetism correction to the $^{12}$B $\beta$-spectrum shape. In \cite{Na87},
the $^{12}$B spectrum itself was used to set the calibration. The $^{12}$B
source was produced through the $^{11}$B$(d,\,p)$ reaction, with the 
spectrograph fixed at the same field setting used for the $^8$B  data.
The momentum measurement was in channels corresponding to a position along
the focal plane. Prior studies of the spectrograph \cite{He85} demonstrated
good linearity between momentum and position. The $^{12}$B raw data 
($\sim 1.1\times 10^5$\ counts) were collected in 36 momentum bins.
The data were fitted with the standard allowed $\beta$-decay 
spectrum, along with the known recoil order corrections (see \cite{Ca90}
and references therein). Both the normalization and the offset in the 
position relative to the radius of curvature were left free in the fit.
The simultaneous minimization of $\chi^2$ with respect to these two 
parameters yielded a good fit $(\chi^2/N_{\rm DF}=1.1)$, and showed no
evidence for a systematic deviation in the residuals. The error
in the absolute momentum calibration was estimated to be less than $0.2\%$
at $1\sigma$. Our conservative estimate for the maximum $(\pm3\sigma)$  
uncertainty in the energy calibration of the $\beta$-decay data in 
\cite{Na87} is $\delta E_\beta = \pm0.090$ MeV. This value corresponds to 
three times a $\pm0.2\%$ error at the endpoint energy, 
$E_{\beta}\simeq 15$ MeV. 

As we will see, the $\beta$-decay spectrum data of \cite{Na87} play 
a fundamental role in constraining the uncertainties of the neutrino 
energy spectrum.   Attempts to confirm the positron spectrum data,
including measurements at lower momenta,
would be  helpful.  Unfortunately, existing  additional  data on 
the $^8$B $\beta$-decay  as reported (only graphically) in 
$\beta$-$\alpha$ correlation searches \cite{Tr75,Ke80}  are too sparse 
to be useful for our purposes.

\subsection{Alpha decay data}

Figure~2 shows our compilation of the experimental data for the delayed 
$\alpha$-spectrum. The  measurements with the highest statistics have 
been performed by Wilkinson and Alburger \cite{Wi71}, using both a thick 
and a thin catcher (WA1 and WA2 in Fig.~2, respectively). For our purposes, 
it is sufficient to know that the typical energy loss of $\alpha$-particles 
is $\sim100$ ($\sim50$) keV in the thick (thin) catcher. The  WA1 data  
($\sim2.1\times 10^6$ counts) have been reported by Warburton \cite{Wa86}, 
together with the proper channel-energy calibration formula. 
Barker  \cite{Ba89} has described similarly the WA2 data  
($\sim2.5\times10^6$ counts). An older measurement of the $\alpha$ 
spectrum was performed by Farmer and Class \cite{Fa60}, although reported 
only in a graphical form (Fig.~2 in \cite{Fa60}). The spectrum labeled 
``FC'' in Fig.~2 corresponds to a digitized form of their data 
($\sim0.5\times10^6$ counts). A high-statistics data set has been also 
collected  by one of us (L.DeB.) and D.~Wright, in the course of a recent 
experimental search for CVC violation and SCC effects in the $A=8$ multiplet 
\cite{Br95}. These data ($\sim1.6\times10^6$ counts) are labeled as ``DBW'' 
in Fig.~2. 

In all of the  $\alpha$-decay  measurements we use, the experimentalists
devoted considerable attention to the $\alpha$-energy calibration, and in 
particular to the accurate estimate of the energy loss in the target. The 
spectra were corrected by the experimentalists for calibration and energy 
loss effects. We do not use  the $\alpha$-spectrum measured  by Clark 
{\em et al.\/} in \cite{Cl69}, since data cannot be extracted from their 
Fig.~2 with sufficient precision to be useful for our purposes.

The spectra in Fig.~2 are peaked at $E_{peak}\simeq1.5$ MeV and decrease
very rapidly for $E_\alpha\neq E_{peak}$ (notice the logarithmic scale).
At $1.5^{+1.8}_{-0.7}$ MeV, the spectral values are decreased
by a factor $\sim 10$. The interval $1.5^{+1.8}_{-0.7}$ MeV contains
$\sim 90\%$ of the experimental counts for each data set. Therefore
the ``tails'' of the spectra ($E_\alpha\lesssim 0.8$ MeV and
$E_\alpha\gtrsim 3.3$ MeV contribute only $\sim 10\%$ to the smearing
of the endpoint energies in the calculation of the neutrino or positron
spectra from $^8$B decay.

\subsection{Discussion of experimental uncertainties}

A close inspection of Fig.~2 reveals  that the four different 
$\alpha$-decay spectra WA1, WA2, FC and DBW have slightly  displaced peaks, 
with a total spread in the peak energies of about  $\pm0.08$ MeV. Indeed, 
the largest uncertainty in these spectra can be ascribed to a possible
bias $b$ in the measured alpha-particle energy: 
$E_\alpha\rightarrow E_\alpha+b$. If the intermediate $2^+$ state at 
3.04 MeV (see Fig.~1) were a narrow resonance, then $b$ could be 
taken as a constant 
bias. In general, $b$ may depend on $E_\alpha$, since in fact the 
intermediate state is not very narrow; thus the bias might assume 
slightly different values at the peak or in the tails of the 
$\alpha$-decay spectra. However, in the calculation of the $^8$B
neutrino spectrum the $\alpha$-spectrum tails are much less important
than the peak region ($0.8 \lesssim E_\alpha \lesssim 3.3$ MeV),
so that $b$ is assumed to be a constant
$(b\simeq b_{peak})$ for our purposes. 

For the spectra WA1 and WA2, Barker has given in \cite{Ba89} a thorough 
discussion and an estimate of the possible contributions to $b$, including 
uncertainties due to energy loss, calibration, and finite resolution.
A succint summary of Barker's analysis is that $|b|\lesssim0.05$ MeV at 
$1\sigma$.  In particular, for the spectrum WA2 (thin catcher), two possible
channel-energy calibration formulae are presented in \cite{Ba89}
[see Eq.~(1) and (15) therein], which differ by $\sim0.05$ MeV in the 
alpha-energy peak. The first  calibration [Eq.~(1) in Barker's paper]  
has been  used in connection with the spectrum WA2 shown in Fig.~2; 
the second  calibration will be used below for a further check of the 
sensitivity of the inferred spectrum to possible systematic uncertainties.
For the DBW spectrum, the uncertainty in $E_\alpha$ is estimated to be of
comparable magnitude ($\sim 0.04$ MeV) \cite{Br95}.
The uncertainty is larger ($|b|\lesssim0.1$ MeV) for the older FC spectrum 
\cite{Fa60}, in part as a result of the  necessity
of converting to graphical data.

Each of the four $\alpha$-particle data sets can be used to estimate the 
theoretical positron spectrum in $^8$B$(\beta^+){}^8$Be decay.
The ingredients of this calculation are the same as for the neutrino
spectrum, except for the radiative corrections. In fact, in beta-decay
the radiative corrections take a different form according to whether  
the $\beta$-particle, or the neutrino, is detected. In the first case,
they have been computed by Sirlin \cite{Si67}. In the second case, they
have been recently evaluated by Batkin and Sundaresan \cite{Su95}, and
will be discussed later (in Sec.~III). The computed positron spectrum,
including radiative corrections, is then compared to the experimental 
$\beta$-decay data (33 points) as in \cite{Na87}. The total number of 
counts collected ($\sim 0.3\times10^6$) is used for normalization,
reducing the number of degrees of freedom ($N_{DF}$) to 32. A normalized
$\chi^2_N$ ($\chi^2_N=\chi^2/N_{DF}$) is then calculated for each input 
$\alpha$ spectrum, and this exercise is repeated also by shifting the 
experimental $\alpha$-energy values ($E_\alpha\rightarrow E_\alpha+b$).
In the calculation of $\chi^2_N$ we include the statistical
errors of the $\beta$-decay data but exclude, in first approximation, their 
energy calibration uncertainty $\delta E_\beta$ (Sec.~II.A). The effect 
of this additional uncertainty is discussed at the end of this section.

Figure~3 shows, for each of the four measured $\alpha$-particle spectra,
the normalized $\chi^2$-fit to the measured positron spectrum as a function 
of the assumed $\alpha$-particle energy bias. For zero energy bias 
(vertical dashed line) the  FC, DBW, WA2, and WA1 data provide increasingly
good fits to the beta-decay spectrum. When an allowance for a non-zero 
bias $b$ is made,  all the four alpha-spectra provide equally good fits 
($\chi^2_N\simeq1$), modulo the shift $E_\alpha\rightarrow E_\alpha+b$. 
The dotted curve, which also provides a good fit for zero bias, corresponds 
to the spectrum WA2 using the second Barker's calibration \cite{Ba89}.
The main difference among the four $\alpha$-particle data sets can thus be 
ascribed to small biases in the measured $\alpha$-energies.

The natural choice for the optimal $\alpha$-particle energy spectrum
is seen to be, in Fig.~3, WA1 with $b=+0.025$ MeV. This is our choice
for the  experimental input data in the calculation of the ``best'' 
neutrino spectrum. An additional variation of $b$ equal to $\pm0.035$ MeV 
produces a $\Delta\chi^2=9$ 
increase in the fit to the $\beta^+$ spectrum, 
defining a $\pm3\sigma$ range for $b$. 

We have studied the sensitivity of the $\chi^2$-fit to the low-statistics
bins by excluding up to 10 bins in the high-energy part of the 
positron spectrum; the highest-energy bins represent $4.5\%$ of the 
experimental counts. The central values of the bias $b$ in these fits 
(excluding some or all of the 10 highest-energy bins) are spread by 
$\pm 0.005$ MeV around $b=0.025$ MeV, with a $3\sigma$ error that can be 
as large as 0.047 MeV when all the 10 bins are excluded. 

The Kolmogorov-Smirnov (K-S) test provides a non-parametric
(bin-free) way of determining the goodness of fit of two distributions.
We have therefore applied a K-S test to the best-fit (unbinned) normalized 
spectrum. We obtain a $3\sigma$ error of $\pm0.056$ MeV for the bias $b$. 
We adopt this error, $\pm0.056$ MeV, as a conservative but relevant 
$3\sigma$ estimate.

Figure~4 shows the experimental beta-decay spectrum \cite{Na87} (dots 
with $1\sigma$ statistical error bars), the best-fit theoretical spectrum 
(solid curve, obtained by using WA1 data with $b=0.025$ MeV), and the 
``$\pm3\sigma$'' theoretical spectra (dotted curves, WA1 data with 
$b=0.025\pm0.056$ MeV). 

As discussed in Sec.~II.A, the $\beta$-decay reference spectrum is 
affected by a maximum  energy calibration uncertainty
$\delta E_\beta=\pm 0.090$ MeV (3$\sigma$). An error $\pm \delta E_\beta$
corresponds to an error $\mp 2\delta E_\alpha$ 
in the $^8$B$(\beta^+){}^8$Be$(2\alpha)$ decay chain. As a consequence,
the total range of the $\alpha$-energy bias $b$ gets slightly enlarged: 
$b=0.025\pm0.056\pm0.045=0.025\pm0.072$ MeV, 
where the two (independent) errors have been added in quadrature.

The estimated  $\pm3\sigma$ 
range for $b$  ($\pm0.072$ MeV) is approximately equal to the 
spread in the peaks of the experimental alpha-decay spectra  ($\pm0.08$ MeV) 
discussed at the beginning of this section. 

\section{The $^8$B standard neutrino spectrum and its uncertainties}

In this section, the main results of the present paper are presented:
a best (standard) $^8$B neutrino spectrum $\lambda(E_\nu)$, together 
with two supplementary spectra, $\lambda^+(E_\nu)$ and $\lambda^-(E_\nu)$,
obtained by stretching the total uncertainties to their $\pm3\sigma$ 
limits. These spectra are given both in figures and in tables.
The effects of the individual experimental and theoretical uncertainties 
are  also discussed and illustrated graphically.

\subsection{Optimal neutrino spectrum and its $3\sigma$ deviations}

The successive calculations of the $^8$B normalized neutrino spectrum 
$\lambda(E_\nu)$  included---besides the phase space factor 
\cite{Ba64}---the  intermediate state smearing \cite{Ba64}, the proper 
Fermi function \cite{Ba64,Ba78}, and the forbidden corrections to 
the allowed transition \cite{Ba86}. 

Napolitano {\em et al.\/} \cite{Na87} pointed out the potential relevance
of radiative corrections, although only those corresponding to the 
$^8$B {\em positron\/} spectrum \cite{Si67} were known at the time.
Here we include the appropriate radiative corrections to the $^8$B 
{\em neutrino\/}  spectrum that have been recently calculated in 
\cite{Su95}. These corrections are smaller (due to a cancellation between 
real and virtual photon contributions \cite{Su95}), and with a milder 
energy dependence, than the corrections that apply when the charged 
lepton is detected in $\beta$-decay \cite{Si67}. As we shall see below, 
their inclusion makes no significant difference in the calculation.

The optimal input alpha-decay data, as discussed in the previous 
section, are taken as WA1 with an energy  bias $b$ having a central value
of $0.025$ MeV and a $\pm3\sigma$ experimental uncertainty 
$\Delta b=\pm0.072$ MeV (determined by the fit to the $\beta$-decay data). 
We anticipate (see below) that the inclusion of the maximal theoretical 
uncertainties can be mimicked by enlarging the above range to 
$\Delta b=\pm0.104$ MeV.  The ``$3\sigma$ different'' neutrino spectra, 
$\lambda^+$ and $\lambda^-$, are  calculated for the  extreme values of 
$\Delta b$ ($+0.104$ and $-0.104$ MeV, respectively).

The results for the normalized neutrino spectra are shown in Fig.~5. 
(The spectra all happen to be almost coincident at about half of the 
maximum energy.) Numerical values of $\lambda$, $\lambda^+$ and 
$\lambda^-$ are reported in Table~I. A computer-readable version
of Table~I is available at the WWW site: 
http://www.sns.ias.edu/\~{}jnb/neutrino.html.

An alternative representation of the above neutrino spectra, in which the 
trivial part of the energy dependence is factorized out, is shown in  
Fig.~6 (a Kurie plot 
\footnote{We have not divided the spectrum by the Fermi function as is 
usual in plotting beta-decay spectra. In the present case, the Fermi 
function would have to be averaged over a range of positron energies 
because of the width of the final ($2^+$) state in $^8$Be.}),
where the ordinate is $\sqrt{\lambda}/E_\nu$. Notice the deviation from 
a  straight line; the deviation is primarily due to the smearing over 
the intermediate broad state of $^8$Be.

A representation of the integral spectrum is shown in Fig.~7, where the
fraction $f$ of $^8$B neutrinos produced above a given neutrino energy 
threshold $E_{th}$  [$f=\int_{E_{th}}^\infty dE_\nu \lambda(E_\nu)$] is 
plotted as function of $E_{th}$.
 
\subsection{Spectrum uncertainties: Experimental and theoretical components}

The effect of varying the input alpha-decay data with respect to the 
optimal choice (WA1 with bias $b=0.025$) is shown in Fig.~8,
where the solid line represents the standard spectrum $\lambda$, and
crosses are placed at representative points along the spectra 
$\lambda^{'}$ obtained with WA1, WA2, FC and DBW input data (with no bias).
The differences $\Delta\lambda=\lambda^{'}-\lambda$ are very small, and 
can be best appreciated in an expanded scale in Fig.~9, where the 
dimensionless quantity  $\Delta\lambda/\lambda_{peak}$ is plotted 
[$\lambda_{peak}=\max{\lambda(E_\nu)}$].

Figure 9  also shows the  $\pm 3\sigma$ deviations 
$(\lambda^\pm-\lambda)/\lambda_{peak}$. These deviations are similar to 
sinusoidal curves, with a maximum amplitude of $\sim2.5\%$. The average 
value of the absolute deviation is thus 
$\langle |\Delta\lambda|/\lambda_{peak}\rangle\simeq
(2/\pi)|\Delta\lambda|_{max}/\lambda_{peak}\simeq1.6\%$ at $3\sigma$. The 
difference between  the Bahcall and Holstein \cite{Ba86} spectrum
$\lambda_{\rm BH}$ and the best-fit spectrum $\lambda$ can also be 
represented well by a sinusoidal curve (like those shown in Fig.~9). The 
amplitude of the difference $\lambda_{\rm BH}-\lambda$ is $\sim0.7\sigma$,
to be compared with the effective $3\sigma$ differences 
$\lambda^\pm-\lambda$ shown in Fig.~9.
Similarly, the difference between the spectrum $\lambda_{\rm N}$
calculated by J.~Napolitano {\em et al.\/} in \cite{Na87} and
the standard spectrum $\lambda$ is about $1.4\sigma$.

Notice that the curve FC in Fig.~9 is somewhat irregular and
``out of phase'' (about $\frac{1}{4}$ of a semiperiod) with respect to
the $\pm3\sigma$ sinusoids. The irregularity is due to the scatter
of the FC data points. The dephasing can be traced back to the fact
that the FC data set has very few low-energy counts,
being limited to $E_\alpha\gtrsim1.2$ MeV. We have
verified that the curve FC in Fig.~9 is ``rephased'' if the FC alpha
spectrum of Fig.~2 is artificially prolonged to lower energies. We have
also computed neutrino spectra with ``mixed alpha data,'' namely with WA1
data for $E_\alpha>E_{peak}$ matched to DBW data for   $E_\alpha<E_{peak}$
(and viceversa). We found that the spectral differences can always
be reabsorbed in a uniform bias $b$ to a very good approximation,
with residual differences of $\lesssim 0.015$ MeV with respect to
the ``unmixed'' original data.
We conclude that the low-energy part of the experimental alpha-decay
spectrum (that affects particularly the high-energy tail of the
neutrino spectrum) is sufficiently well known for our purposes, and
that the alpha spectra uncertainties can be effectively parametrized as
a uniform energy offset $b$.

The effects of radical assumptions about the correctness of the 
theoretical calculations beyond the (phase space)$\times$(Fermi function)
approximation are shown in Fig.~10, on the same scale as  Fig.~9. The 
curves labeled 1 and 2 represent the shifts $\Delta\lambda/\lambda_{peak}$ 
obtained by setting to zero the forbidden or the radiative corrections, 
respectively.  For curve 3, the radiative corrections were 
(inappropriately)  assumed to be the same as for the positron detection
case \cite{Si67} (as was also done in \cite{Na87}); the present exercise 
is intended to account roughly for a  hypothetical situation  in which 
the cancellation between real and virtual photon contribution might not 
be as effective as computed in \cite{Su95}. 

The maximum theoretical deviation (curve 1 in Fig.~10) is obtained by 
excluding the forbidden corrections altogether. We have  verified that 
this deviation can be mimicked by recalculating the neutrino energy 
spectrum with an additional energy bias, $\Delta b_{\rm theo}\simeq0.075$ 
MeV. This value is comparable to  the estimated $3\sigma$ experimental 
uncertainty evaluated in Sec.~II~C, $\Delta b_{\rm exp}\simeq0.072$. We 
assume that the maximum theoretical offset, $\Delta b_{\rm theo}$, 
corresponds to an ``effective'' $3\sigma$ statistical significance. Our 
final  best-estimate for the bias $b$ to be applied to the reference WA1
$\alpha$-decay spectrum is therefore: $b=0.025 {\rm\  MeV}
\pm(\Delta b^2_{\rm exp}+\Delta b^2_{\rm theo})^\frac{1}{2}=0.025\pm0.104$
MeV, as anticipated in Sec.~III.A. 

The assignment of a ``$3\sigma$ level of significance'' to the offset that
parameterizes the total effect of the forbidden corrections is a plausible
estimate. However, there is no rigorous way to estimate a $3\sigma$ 
theoretical uncertainty. Our estimate is motivated by the fact that the 
calculation of  forbidden corrections in beta-decay is not made purely from 
first principles (strong and electroweak lagrangian), unlike the QED 
radiative corrections. The evaluation of the forbidden terms makes use of 
approximate symmetries to expand the weak nuclear current in terms of a few 
nuclear form factors, that are evaluated using  nuclear models and
difficult $\beta$-$\alpha$ correlation experiments (see \cite{Ba86} and
references therein).

If the reader prefers to adopt a smaller or larger estimate for the 
theoretical uncertainty, a simple prescription for estimating the changes 
in the inferred uncertainties is given below.  The prescription is based 
upon linear error propagation and generally gives agreement with an exact 
numerical calculation to a fractional accuracy of about 5\% or better 
in the change induced by the rescaling of the error.
Recalculate the total $3\sigma$ range for the bias, 
$\Delta b=(\Delta b^2_{\rm exp}+\Delta b^2_{\rm theo})^\frac{1}{2}$,
with $\Delta b_{\rm exp}=0.072$ MeV and the reader's preferred estimate for 
$\Delta b_{\rm theo}$. Then rescale by a factor 
$f=\Delta b/(0.104{\rm\  MeV})$  all the  $^8$B-related $3\sigma$
total uncertainties quoted in this paper. For example, the $3\sigma$ 
spectral deviations from the best-estimated neutrino spectrum shape, 
$\lambda^\pm(E_\nu)-\lambda(E_\nu)$, become 
$f[\lambda^\pm(E_\nu)-\lambda(E_\nu)]$. Notice that the rescaling 
factor is at least \mbox{$f=0.69$}, which is obtained by setting 
$\Delta b_{\rm theo}=0$. 

A final remark is in order. The relative contribution of the different 
$2^+$ states of $^8$Be in the  $\alpha$-decay spectrum 
(see Fig.~1) has been analyzed  by Barker 
\cite{Ba69,Ba89} and Warburton \cite{Wa86} within the $R$-matrix 
formalism. Their results are not in complete agreement, the fitted 
amplitude of the intermediate states being  sensitive to the input data 
(see, e.g., the discussion in \cite{Ba89}). In particular, $R$-matrix fits
are very sensitive to the absolute energy calibration, as well as to the
tails of the alpha-decay spectrum. In our calculation of the 
neutrino spectrum, the absolute $\alpha$-energy is allowed to vary
within the quoted uncertainties ($\pm 0.104$ MeV). The
tails of the alpha decay spectrum are not decisive for our purposes.
We are not really interested in separating the relative
contributions of the $2^+$ states; we just rely upon the global 
population of the $2^+$ states as derived from
the {\em experimental\/} $\alpha$-decay data.
We conclude that the disagreement between  the theoretical $R$-matrix
fits  \cite{Ba69,Wa86,Ba89} is not an issue in our calculation of
the $^8$B neutrino spectrum.

\section{Absorption cross section of $^8$B neutrinos in 
$^{37}$C\lowercase{l} and $^{71}$G\lowercase{a} }

In this section, we present improved calculations of the $^8$B neutrino 
absorption cross section for chlorine and for gallium. Recent calculations 
of the chlorine absorption cross section were made by Bahcall and Holstein 
\cite{Ba86}, Garc{\'\i}a {\em et al.\/} \cite{Ga91}, and Aufderheide  
{\em et al.\/} \cite{Au94}; the results of earlier calculations
can be found in  \cite{Ba64,Ba78,Ba00,Ba66,Se74,Ha77,Ra81}. The most 
recent previous calculation of the gallium cross section $\sigma_{\rm Ga}$ 
was  by  Bahcall and Ulrich \cite{Ba88}.

Transitions to excited states dominate the total cross section in either 
of the absorption processes $^{37}{\rm Cl}(\nu,\,e){}^{37}{\rm Ar}$ and
$^{71}{\rm Ga}(\nu,\,e){}^{71}{\rm Ge}$.  The Gamow--Teller transition
strengths, $B$(GT), can be estimated from the rates of the analogous
 charge-exchange $(p,\,n)$ reactions. For the $A=37$ system, these 
transition matrix elements can  be determined experimentally by studying 
the $^{37}{\rm Ca}(\beta^+){}^{37}{\rm K}$ transition \cite{Ba00,Ba64}, 
which is the isospin mirror process of 
$^{37}{\rm Cl}(\nu,\,e){}^{37}{\rm Ar}$.  The interested reader is referred 
to the recent review in \cite{Ku94} for a more extensive discussion 
of these processes.

\subsection{Absorption Cross Section for Chlorine}

Including for the first time forbidden corrections, Bahcall and Holstein 
\cite{Ba86} calculated  the $^8$B neutrino cross section on chlorine and 
obtained:  
\begin{equation} 
\sigma_{\rm Cl}=(1.06\pm 0.10)\times10^{-42}{\rm \ \ cm}^2\,.
\end{equation}
The quoted uncertainties represented the maximum estimated error 
$(3\sigma)$. The calculation made use of the $B$(GT) values derived from 
the $^{37}$Ca $\beta$-decay, as reported by Sextro  {\em et al.\/} in 
\cite{Se74}. The estimated $3\sigma$ error $(\pm0.10)$ had two components, 
$\pm0.08$ from $^8$Be $\alpha$-decay data and $\pm0.06$ from $^{37}$Ca 
$\beta$-decay data uncertainties, to be added in quadrature.

Using the same  low-energy data \cite{Se74} as in \cite{Ba86}, and the 
spectra $\lambda$, $\lambda^+$, and $\lambda^-$ reported in Table~I, we 
find $\sigma_{\rm Cl}=(1.08\pm 0.15)\times10^{-42}$ cm$^2$. The best-estimate 
calculated cross sections differ by $2\%$. The $3\sigma$ error component 
from $^8$Be$(2\alpha)$ decay data is $\pm0.07$.

Recently, new experiments have been carried out 
to determine  more precisely the $B$(GT) strengths in 
$^{37}{\rm Cl}(p,\,n){}^{37}{\rm Ar}$ \cite{Ra81} and
$^{37}{\rm Ca}(\beta^+){}^{37}{\rm K}$ \cite{Ga91} processes. Taken at 
face values, the  $B$(GT) strengths  derived by the different experiments 
\cite{Ga91,Se74,Ra81} were not in good agreement. Critical examinations 
\cite{Crit} of the data analyses, as well as supplementary data 
\cite{Il93}, may have led to a satisfactory understanding \cite{Au94} of the 
low-energy levels and their $B$(GT) strengths in the  $A=37$ system. 

Using the latest available data \cite{Au94} and the neutrino spectra 
$\lambda^{(\pm)}$ calculated here, we find
\begin{equation}
\sigma_{\rm Cl}=(1.14\pm0.11)\times10^{-42}{\rm \ \ cm}^2\,. 
\end{equation}
Equation (2) represents our best estimate, and the associated $3\sigma$ 
uncertainties, for the $^8$B neutrino absorption cross section on chlorine.
The contribution to the total error from the measured $B$(GT) values is  
assumed to be $\pm0.08$, as in the analysis \cite{Au94}. The difference 
between the  values of the chlorine absorption cross section in Eq.~(1) 
and Eq.~(2) is $7\%$, well within the quoted errors.

Figure~11 shows the energy dependence of our our best-estimate chlorine 
cross section (solid line).  Values of $\sigma_{\rm Cl}$ for representative 
neutrino energies are also given in Table~II. The difference between the 
present and  the previous calculation of the {\em energy-dependent\/} 
cross section by Bahcall and Ulrich \cite{Ba88} (dashed line in Fig.~11) 
is  less than $20\%$ for $E_\nu<16$ MeV. The differences are largest at 
the highest energies since the newer data include transitions to higher
excitation states in $^{37}$Ar that were not determined in the previous
experiments (see \cite{Ga91,Crit,Au94}).

To give the reader some perspective on how the $^8$B neutrino spectrum 
and the chlorine absorption cross section have changed with time, we give 
in Table~III all the published values of $\sigma_{\rm Cl}$ with which we 
are familiar. The calculated cross sections have been approximately 
constant, within the estimated errors, since 1978, although there have 
been numerous refinements (which are described in 
\cite{Ba78,Ba86,Ra81,Ga91,Au94}). The reasons for the relatively 
significant change in the 1978 best-estimated value \cite{Ba78}
with respect to the earlier calculations \cite{Ba64,Ba66,Se74,Ha77}
are described in the last paragraph of Sec.~IV.B.3 in \cite{Ba78}.

\subsection{Absorption Cross Section for Gallium}

The $^8$B neutrino absorption  cross section for gallium that is widely 
used was calculated by Bahcall and Ulrich \cite{Ba88} and is:
\begin{equation}
\sigma_{\rm Ga}=(2.43^{+2.1}_{-1.1})\times10^{-42}{\rm \ \ cm}^2\,,
\end{equation}
where the quoted uncertainties represented the maximum estimated  errors 
$(3\sigma)$. The $B$(GT) values used in  the quoted calculation were taken 
from a  $^{71}{\rm Ga}(p,\,n){}^{71}{\rm Ge}$ experiment performed by 
Krofcheck {\em et al.\/} \cite{Kr85}. 

The only important recently-published experimental development with which 
we are familiar is the  recent $^{51}$Cr source experiment for the GALLEX 
detector \cite{CrEx}. Hata and Haxton  \cite{Ha95} have shown that the 
measurements  with the chromium source are consistent with the $B$(GT) 
values for the first two excited states that were inferred by Krofcheck 
{\em et al.\/} \cite{Kr85}. 

Therefore we repeat the Bahcall--Ulrich calculation [Eq.~(3)] using the 
best $^8$B neutrino spectrum from the present paper. We find:
\begin{equation}
\sigma_{\rm Ga}=(2.46^{+2.1}_{-1.1})\times10^{-42}{\rm \ \ cm}^2\,.
\end{equation}
The change in the best-estimate cross section is only $\sim1\%$
[relative to Eq.~(3)], which is much smaller than the guessed systematic 
errors, that represent uncertainties in the  interpretation of the 
$(p,\,n)$ measurements.


\section{Summary}

In the previous sections, the spectrum of neutrinos produced in the 
$^8$B$(\beta^+){}^8$Be$(2\alpha)$ decay has been computed, using  
state-of-the-art theory of beta-decay. The laboratory data on the 
associated positron spectrum have been used to  choose an optimal data 
set among the different measured  $^8$Be$(2\alpha)$ decay spectra. 
The experimental and theoretical uncertainties can  both be represented 
well as an  energy shift $(b)$ in the $\alpha$-decay data. The total 
$\pm3\sigma$ range for this shift (bias) has been  conservatively 
estimated to be $\pm 0.104$ MeV. 
A best-fitting standard spectrum,~$\lambda$,~has been 
computed, as well as the ``effective  $\pm3\sigma$'' neutrino spectral 
shapes $\lambda^+$~and~$\lambda^-$  [$\lambda^\pm-\lambda=3$ 
standard deviations]. The standard spectrum $\lambda$ differs by about 
$0.7\sigma$ from the Bahcall and Holstein neutrino spectrum 
\cite{Ba86} and by about $1.4\sigma$ from the Napolitano  {\em et al.\/} 
spectrum \cite{Na87}.

The $^8$B neutrino absorption cross section for chlorine calculated
with the best-fitting spectrum derived here  and with the most recent 
data on the low-lying states in the $A=37$ system is
$\sigma_{\rm Cl}=(1.14\pm{0.11})\times10^{-42}$ cm$^2$ $(3\sigma)$.
This result is in agreement with the estimates derived in 1964 
(see Table~III). The best-estimate gallium absorption cross section is 
$\sigma_{\rm Ga}=(2.46^{+2.1}_{-1.1})\times10^{-42}$ cm$^2$ $(3\sigma)$.

Many readers would prefer to quote $1\sigma$ rather than $3\sigma$ errors.
To a good approximation,  $1\sigma$ errors can be obtained from our quoted
$3\sigma$ values by dividing by three. Moreover,
$[\lambda^\pm(E_\nu)-\lambda(E_\nu)]_{1\sigma}\simeq
 [\lambda^\pm(E_\nu)-\lambda(E_\nu)]_{3\sigma}/3$.

All the available experimental data  on the 
$^8$B$(\beta^+){}^8$Be$(2\alpha)$ decay are consistent  within
the quoted uncertainties. Higher order contributions in the theoretical 
calculation of the $^8$B neutrino spectrum should  be very small.
Any measured deviation of the $^8$B solar neutrino spectrum in excess of
the conservative limits given in this paper could be considered as evidence
for  new physics beyond the standard electroweak model, at an effective 
significance level greater than three standard deviations.

\acknowledgments

We acknowledge useful discussion and correspondence with 
F.\ Ajzenberg-Selove, M.~B.\ Aufderheide, I.~S.\ Batkin, A.\ Hime, 
and M.~K.\ Sundaresan.  We are grateful to D.~H.\ Wilkinson 
for permission to use the data in \cite{Wi71} and for valuable discussions.
We thank the anonymous referee for very useful suggestions.
The work of J.N.B.\ was supported in part by NSF grant No.~PHY92-45317.
The work of E.L.\ was supported in part by the  Institute for Advanced 
Study through the NSF grant listed above and through the Monell Foundation,
and in part by INFN. The research of E.L.\ was also performed under the 
auspices of the Theoretical Astroparticle Network, under contract 
No.\ CHRX-CT93-0120 of the Direction General XII of the E.E.C. 
The work of D.E.A. at Brookhaven National Laboratory was performed under 
contract No.\ DE-AC02-76CH00016 with the U.S.\ Department of Energy.
The work of S.J.F.\ was performed  under 
contract No.\ DE-AC03-76SF00098 with the U.S.\ Department of Energy.
 

\begin{table}
\caption{The spectrum $\lambda$ of solar neutrinos from the decay of $^8$B,
together with the spectra $\lambda^\pm$ associated with the total $\pm3\sigma$
uncertainties. The neutrino energy $E_\nu$ is MeV and $\lambda^{(\pm)}(E_\nu)$
is the probability that a neutrino with energy $E_\nu$ is emitted between 
$E_\nu\pm0.05$ MeV. A computer-readable version of this table is available 
at the WWW site:  http://www.sns.ias.edu/\~{}jnb/neutrino.html.}

\begin{tabular}{rccccrccc}
$E_\nu$ & $\lambda(E_\nu)$ & $\lambda^+(E_\nu)$ & $\lambda^-(E_\nu)$
&&
$E_\nu$ & $\lambda(E_\nu)$ & $\lambda^+(E_\nu)$ & $\lambda^-(E_\nu)$ 
\\
\hline
  0.1 & .000202 & .000228 & .000168 &&   8.1 & .118422 & .117881 & .118768
 \\
  0.2 & .000742 & .000800 & .000636 &&   8.2 & .116796 & .116108 & .117284
 \\
  0.3 & .001522 & .001619 & .001376 &&   8.3 & .115094 & .114259 & .115722
 \\
  0.4 & .002516 & .002708 & .002356 &&   8.4 & .113317 & .112337 & .114086
 \\
  0.5 & .003775 & .004058 & .003543 &&   8.5 & .111468 & .110345 & .112378
 \\
  0.6 & .005254 & .005628 & .004924 &&   8.6 & .109552 & .108286 & .110600
 \\
  0.7 & .006938 & .007398 & .006514 &&   8.7 & .107570 & .106164 & .108755
 \\
  0.8 & .008804 & .009357 & .008282 &&   8.8 & .105525 & .103980 & .106847
 \\
  0.9 & .010835 & .011494 & .010219 &&   8.9 & .103421 & .101740 & .104877
 \\
  1.0 & .013020 & .013795 & .012304 &&   9.0 & .101261 & .099446 & .102849
 \\
  1.1 & .015348 & .016242 & .014525 &&   9.1 & .099047 & .097101 & .100765
 \\
  1.2 & .017809 & .018816 & .016871 &&   9.2 & .096784 & .094710 & .098629
 \\
  1.3 & .020386 & .021508 & .019334 &&   9.3 & .094475 & .092276 & .096444
 \\
  1.4 & .023066 & .024308 & .021902 &&   9.4 & .092122 & .089802 & .094213
 \\
  1.5 & .025840 & .027201 & .024564 &&   9.5 & .089731 & .087295 & .091940
 \\
  1.6 & .028696 & .030170 & .027306 &&   9.6 & .087303 & .084756 & .089626
 \\
  1.7 & .031624 & .033211 & .030121 &&   9.7 & .084844 & .082189 & .087277
 \\
  1.8 & .034611 & .036312 & .033001 &&   9.8 & .082355 & .079598 & .084895
 \\
  1.9 & .037650 & .039465 & .035934 &&   9.9 & .079842 & .076987 & .082484
 \\
  2.0 & .040733 & .042659 & .038910 &&  10.0 & .077308 & .074361 & .080047
 \\
  2.1 & .043851 & .045885 & .041923 &&  10.1 & .074758 & .071723 & .077588
 \\
  2.2 & .046996 & .049133 & .044965 &&  10.2 & .072194 & .069078 & .075110
 \\
  2.3 & .050159 & .052398 & .048028 &&  10.3 & .069620 & .066429 & .072617
 \\
  2.4 & .053332 & .055669 & .051106 &&  10.4 & .067041 & .063780 & .070113
 \\
  2.5 & .056509 & .058939 & .054190 &&  10.5 & .064460 & .061137 & .067602
 \\
  2.6 & .059681 & .062202 & .057274 &&  10.6 & .061881 & .058503 & .065087
 \\
  2.7 & .062844 & .065450 & .060352 &&  10.7 & .059309 & .055882 & .062572
 \\
  2.8 & .065989 & .068678 & .063418 &&  10.8 & .056747 & .053279 & .060061
 \\
  2.9 & .069112 & .071877 & .066464 &&  10.9 & .054199 & .050698 & .057557
 \\
  3.0 & .072206 & .075040 & .069486 &&  11.0 & .051670 & .048142 & .055065
 \\
  3.1 & .075265 & .078163 & .072479 &&  11.1 & .049162 & .045617 & .052587
 \\
  3.2 & .078282 & .081238 & .075436 &&  11.2 & .046681 & .043126 & .050129
 \\
  3.3 & .081253 & .084263 & .078353 &&  11.3 & .044231 & .040672 & .047693
 \\
  3.4 & .084173 & .087230 & .081224 &&  11.4 & .041814 & .038262 & .045283
 \\
  3.5 & .087038 & .090136 & .084044 &&  11.5 & .039435 & .035897 & .042904
 \\
  3.6 & .089841 & .092976 & .086809 &&  11.6 & .037097 & .033583 & .040559
 \\
  3.7 & .092580 & .095746 & .089516 &&  11.7 & .034805 & .031322 & .038251
 \\
  3.8 & .095251 & .098441 & .092159 &&  11.8 & .032563 & .029120 & .035984
 \\
  3.9 & .097851 & .101059 & .094736 &&  11.9 & .030373 & .026979 & .033762
 \\
  4.0 & .100374 & .103595 & .097243 &&  12.0 & .028240 & .024904 & .031588
 \\
  4.1 & .102818 & .106045 & .099677 &&  12.1 & .026167 & .022897 & .029466
 \\
  4.2 & .105180 & .108408 & .102034 &&  12.2 & .024158 & .020962 & .027399
 \\
  4.3 & .107457 & .110678 & .104311 &&  12.3 & .022216 & .019103 & .025391
 \\
  4.4 & .109645 & .112855 & .106506 &&  12.4 & .020343 & .017322 & .023445
 \\
  4.5 & .111742 & .114935 & .108617 &&  12.5 & .018544 & .015623 & .021564
 \\
  4.6 & .113747 & .116916 & .110640 &&  12.6 & .016821 & .014007 & .019751
 \\
  4.7 & .115657 & .118796 & .112572 &&  12.7 & .015177 & .012481 & .018009
 \\
  4.8 & .117469 & .120572 & .114414 &&  12.8 & .013614 & .011045 & .016340
 \\
  4.9 & .119181 & .122243 & .116162 &&  12.9 & .012134 & .009700 & .014748
 \\
  5.0 & .120793 & .123808 & .117814 &&  13.0 & .010741 & .008448 & .013235
 \\
  5.1 & .122302 & .125264 & .119369 &&  13.1 & .009435 & .007291 & .011803
 \\
  5.2 & .123707 & .126611 & .120825 &&  13.2 & .008223 & .006231 & .010454
 \\
  5.3 & .125007 & .127846 & .122182 &&  13.3 & .007102 & .005267 & .009190
 \\
  5.4 & .126201 & .128969 & .123438 &&  13.4 & .006073 & .004400 & .008011
 \\
  5.5 & .127287 & .129980 & .124592 &&  13.5 & .005138 & .003630 & .006920
 \\
  5.6 & .128265 & .130876 & .125643 &&  13.6 & .004296 & .002956 & .005922
 \\
  5.7 & .129134 & .131659 & .126590 &&  13.7 & .003548 & .002374 & .005014
 \\
  5.8 & .129893 & .132329 & .127432 &&  13.8 & .002892 & .001880 & .004196
 \\
  5.9 & .130544 & .132885 & .128170 &&  13.9 & .002325 & .001469 & .003468
 \\
  6.0 & .131085 & .133327 & .128802 &&  14.0 & .001843 & .001133 & .002830
 \\
  6.1 & .131517 & .133656 & .129329 &&  14.1 & .001442 & .000863 & .002277
 \\
  6.2 & .131839 & .133870 & .129751 &&  14.2 & .001113 & .000650 & .001807
 \\
  6.3 & .132052 & .133972 & .130069 &&  14.3 & .000849 & .000485 & .001415
 \\
  6.4 & .132158 & .133961 & .130282 &&  14.4 & .000640 & .000358 & .001094
 \\
  6.5 & .132155 & .133839 & .130391 &&  14.5 & .000478 & .000263 & .000835
 \\
  6.6 & .132045 & .133607 & .130395 &&  14.6 & .000354 & .000190 & .000630
 \\
  6.7 & .131830 & .133265 & .130297 &&  14.7 & .000259 & .000137 & .000471
 \\
  6.8 & .131508 & .132815 & .130096 &&  14.8 & .000188 & .000097 & .000349
 \\
  6.9 & .131083 & .132258 & .129795 &&  14.9 & .000135 & .000068 & .000256
 \\
  7.0 & .130555 & .131596 & .129393 &&  15.0 & .000096 & .000047 & .000186
 \\
  7.1 & .129925 & .130829 & .128892 &&  15.1 & .000067 & .000031 & .000134
 \\
  7.2 & .129196 & .129961 & .128294 &&  15.2 & .000047 & .000021 & .000095
 \\
  7.3 & .128368 & .128992 & .127599 &&  15.3 & .000031 & .000013 & .000067
 \\
  7.4 & .127443 & .127925 & .126809 &&  15.4 & .000021 & .000008 & .000046
 \\
  7.5 & .126423 & .126761 & .125927 &&  15.5 & .000014 & .000005 & .000031
 \\
  7.6 & .125310 & .125502 & .124952 &&  15.6 & .000009 & .000003 & .000021
 \\
  7.7 & .124106 & .124152 & .123887 &&  15.7 & .000005 & .000002 & .000014
 \\
  7.8 & .122813 & .122712 & .122734 &&  15.8 & .000003 & .000001 & .000009
 \\
  7.9 & .121433 & .121186 & .121496 &&  15.9 & .000002 & .000000 & .000005
 \\
  8.0 & .119968 & .119574 & .120172 &&  16.0 & .000001 & .000000 & .000003
 \\
\end{tabular}
\end{table}

\vspace*{-11mm}
\begin{table}
\caption{Values of the absorption cross section in chlorine, in units of 
$10^{-42}$ cm$^2$, for representative values of the neutrino  energy. 
The second column refers to the calculation in the present paper. The
third column refers to the Bahcall--Ulrich (BU) \protect\cite{Ba88} 
calculation.}
\begin{tabular}{ccc}
 $E_\nu$ (MeV) & $\sigma_{\rm Cl}$ & \ \ \ $\sigma_{\rm Cl}$(BU) \ \ \ \  \\
\hline
  1	& 5.21E+00 	& 5.21E+00   \\
  2	& 3.70E+01  	& 3.70E+01   \\
  3	& 1.02E+02 	& 1.15E+02   \\
  4	& 2.23E+02 	& 2.63E+02   \\
  5	& 5.38E+02 	& 5.63E+02   \\
  6	& 1.44E+03 	& 1.52E+03   \\
  7	& 4.62E+03 	& 4.76E+03   \\
  8	& 1.01E+04 	& 1.02E+04   \\
  9	& 1.85E+04 	& 1.79E+04   \\
  10	& 3.00E+04 	& 2.77E+04   \\
  11	& 4.45E+04 	& 3.97E+04   \\
  12	& 6.21E+04 	& 5.38E+04   \\
  13	& 8.27E+04  	& 7.00E+04   \\
  14	& 1.06E+05 	& 8.83E+04   \\
  15	& 1.33E+05 	& 1.09E+05   \\
  16	& 1.62E+05 	& 1.31E+05   \\
  18	& 2.28E+05 	& 1.81E+05   \\
  20	& 3.05E+05 	& 2.38E+05   \\
  30	& 8.20E+05 	& 6.11E+05  
\end{tabular}
\end{table}

\begin{table}
\caption{Values of the $^8$B neutrino absorption cross section for chlorine
$(\sigma_{\rm Cl})$, as calculated by various authors. The first and second 
(when given) error components, $\epsilon_{\rm GT}$ and $\epsilon_{\rm B}$, 
are to be added in quadrature; they  refer to the estimated uncertainties 
from the Gamow--Teller (GT) strengths and from the  $^8$B neutrino spectrum, 
respectively.  When the definitions of the errors given in the original 
papers were sufficiently precise, we have indicated (in parentheses) that 
we are quoting $3\sigma$ errors.}
\begin{tabular}{clcl}
\ \ Year & Author(s)		&\ Ref.	
         & $\sigma_{\rm Cl}\pm\epsilon_{\rm GT}\pm\epsilon_{\rm B}$\ 
         ($10^{-42}$ cm$^2$) \\
\hline
\ \ 1964 & Bahcall                    &\ \cite{Ba00} & $1.27\pm0.31$        \\
\ \ 1964 & Bahcall		      &\ \cite{Ba64} & $1.30\pm0.29$        \\
\ \ 1966 & Bahcall		      &\ \cite{Ba66} & $1.35\pm0.10$        \\
\ \ 1974 & Sextro      {\em et al.}   &\ \cite{Se74} & $1.31$               \\
\ \ 1977 & Haxton and Donnelly	      &\ \cite{Ha77} & $1.27\pm0.22\pm0.06$ \\
\ \ 1978 & Bahcall		      &\ \cite{Ba78} & $1.08\pm0.10$        \\
\ \ 1981 & Rapaport    {\em et al.}   &\ \cite{Ra81} & $0.98\pm0.07$        \\
\ \ 1986 & Bahcall and Holstein	      &\ \cite{Ba86} & $1.06\pm0.06\pm0.08$\ 
                                                       $[3\sigma]$          \\
\ \ 1991 & Garc{\'\i}a {\em et al.}   &\ \cite{Ga91} & $1.09\pm0.03$        \\
\ \ 1994 & Aufderheide {\em et al.}   &\ \cite{Au94} & $1.11\pm0.08$\ 
                                                       $[3\sigma]$          \\
\ \ 1996 & Bahcall     {\em et al.}   & This work    & $1.14\pm0.08\pm0.08$\ 
                                                       $[3\sigma]$
\end{tabular}
\end{table}

\begin{figure}
\caption{Energy levels in the $^8$B$(\beta^+){}^8$Be$(2\alpha)$ decay 
chain (not to scale).\hbox{\hskip3.3cm}}
\end{figure}
\begin{figure}
\caption{Compilation of $^8$Be$(2\alpha)$ decay data. The bin widths
are different for different experiments. The data WA1 and WA2 are shifted
on the vertical axis.}
\end{figure}
\begin{figure}
\caption{Values of the normalized
chi-square in a fit to the experimental positron spectrum, using the 
input alpha-decay data of Fig.~2, with an allowance for a possible
bias, $b$, in the detected $\alpha$-particle energy. The  curves are 
remarkably similar, modulo a constant bias.}
\end{figure}
\begin{figure}
\caption{Experimental data on the positron spectrum, together with
the best fit and the $\pm3\sigma$ fits, corresponding
to WA1 alpha-decay data within the bias range $b=0.025\pm0.056$ MeV.}
\end{figure}
\begin{figure}
\caption{The best-estimate (standard) 
$^8$B neutrino spectrum $\lambda$, together with
the spectra $\lambda^\pm$ allowed by the maximum ($\pm3\sigma$)
theoretical and experimental uncertainties.}
\end{figure}
\begin{figure}
\caption{The spectra $\lambda$, $\lambda^+$, and
$\lambda^-$, shown as a Kurie plot.\hbox{\hskip5.33cm}}
\end{figure}
\begin{figure}
\caption{Fraction of $^8$B neutrinos produced with energy $E_\nu$ above
a given threshold $E_{th}$.\hbox{\hskip4.8mm}}
\end{figure}
\begin{figure}
\caption{Variations in the standard neutrino spectrum 
$\lambda$ (solid line) induced
by different input data sets (crosses).}
\end{figure}
\begin{figure}
\caption{Variations $\Delta\lambda$ in the 
standard neutrino spectrum $\lambda$ induced by different
input data sets, divided by the peak value of $\lambda$ ($\lambda_{peak}$).
The maximum $(\pm3\sigma)$ differences $\lambda^\pm-\lambda$ are also shown.}
\end{figure}
\begin{figure}
\caption{Variations $\Delta\lambda$
in the neutrino spectrumy induced by drastic changes
in the theoretical computation, shown in the same scale as  Fig.~9.
See the text for details.}
\end{figure}
\begin{figure}
\caption{Absorption cross section in chlorine (solid line) as a function
of the neutrino energy. The dashed line refers to the
Bahcall--Ulrich \protect\cite{Ba88} calculation.}
\end{figure}
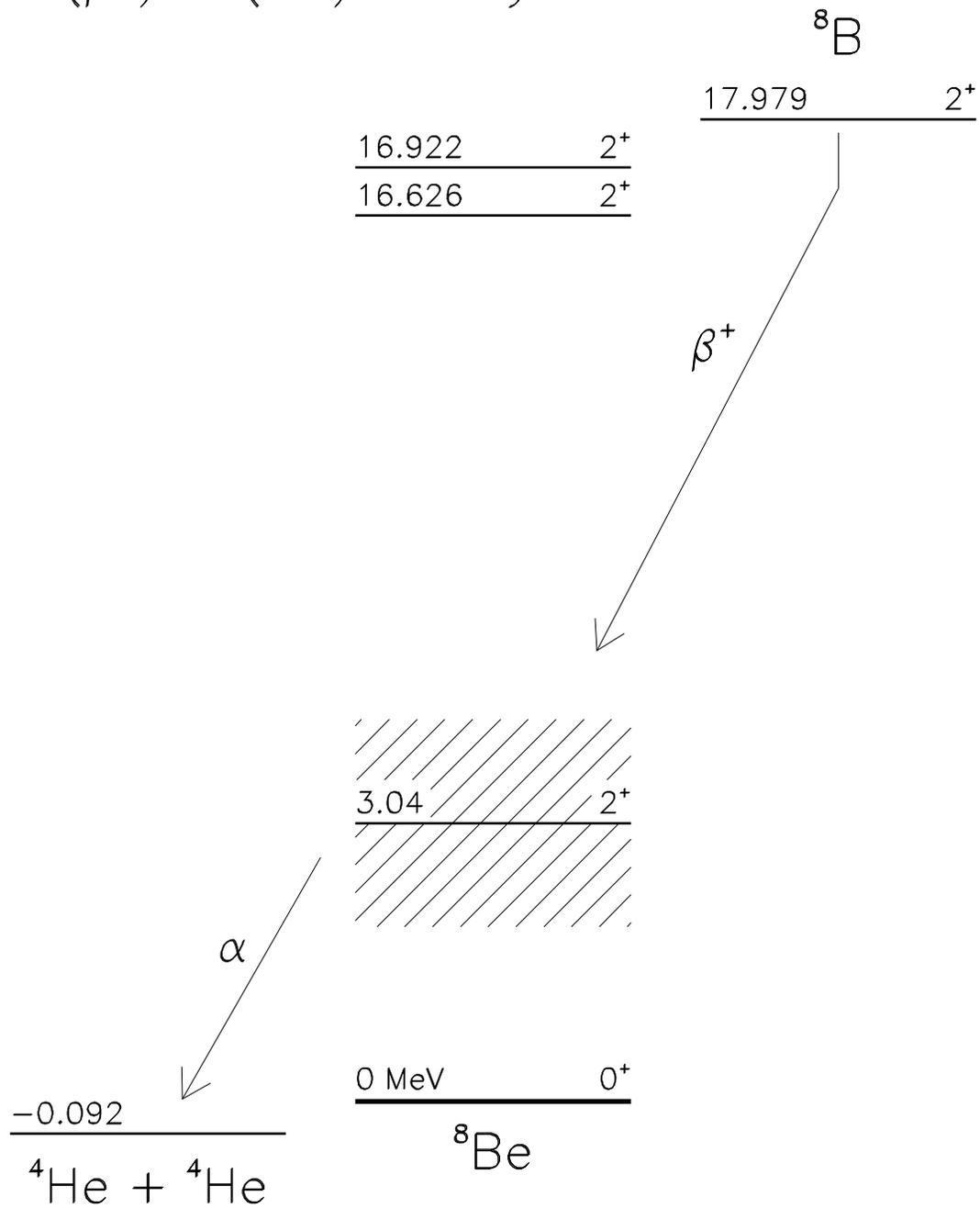
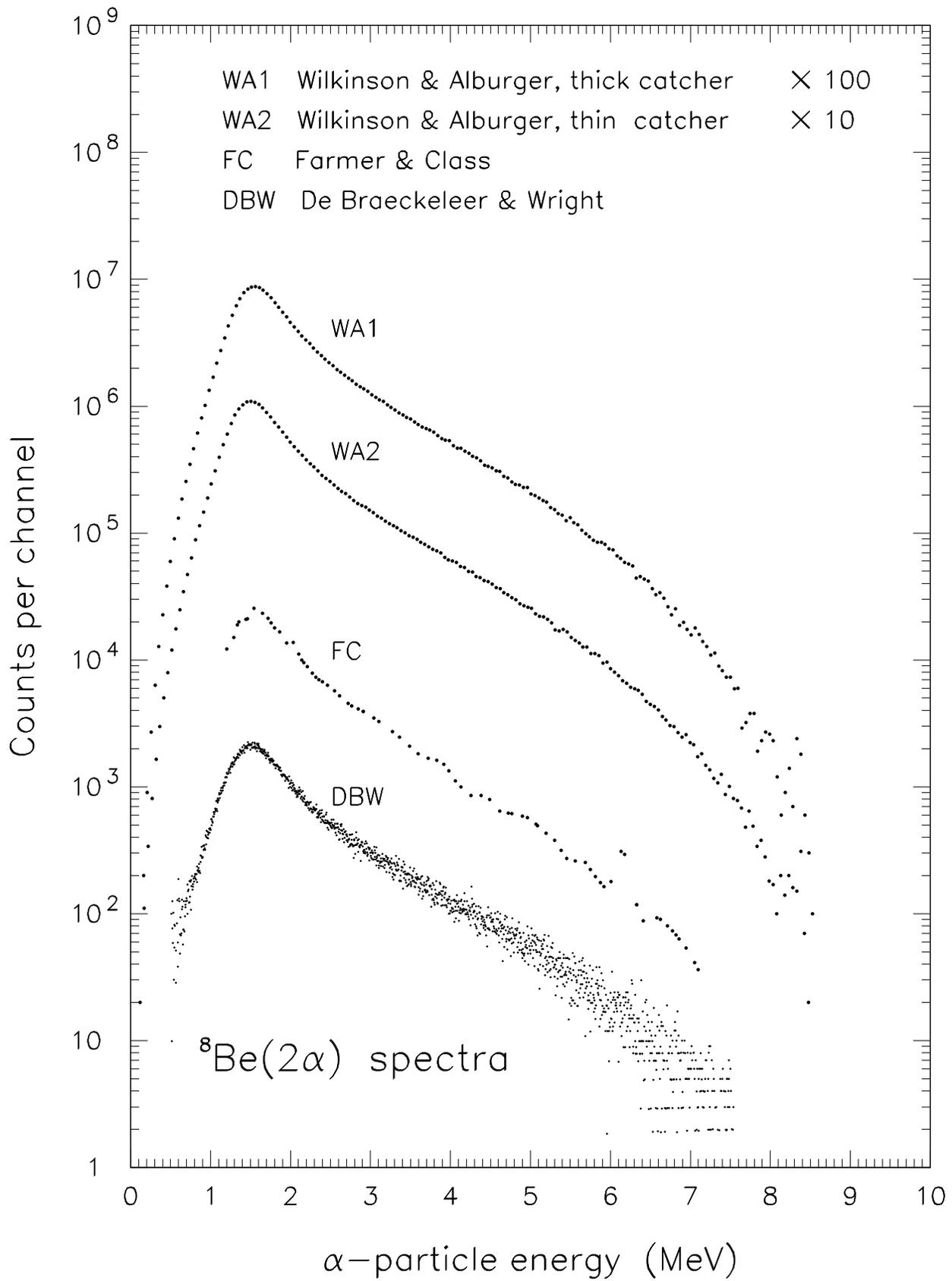
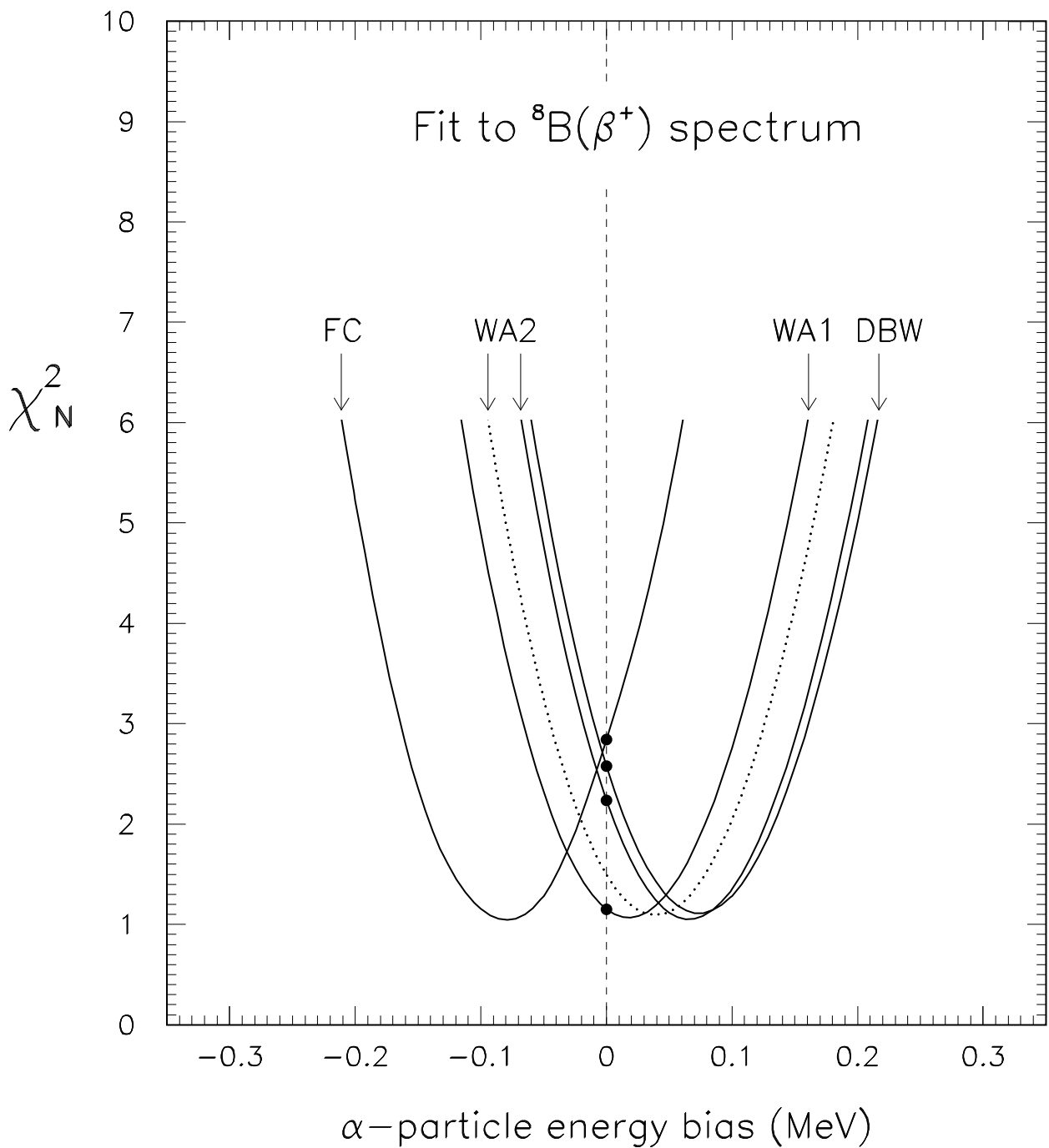
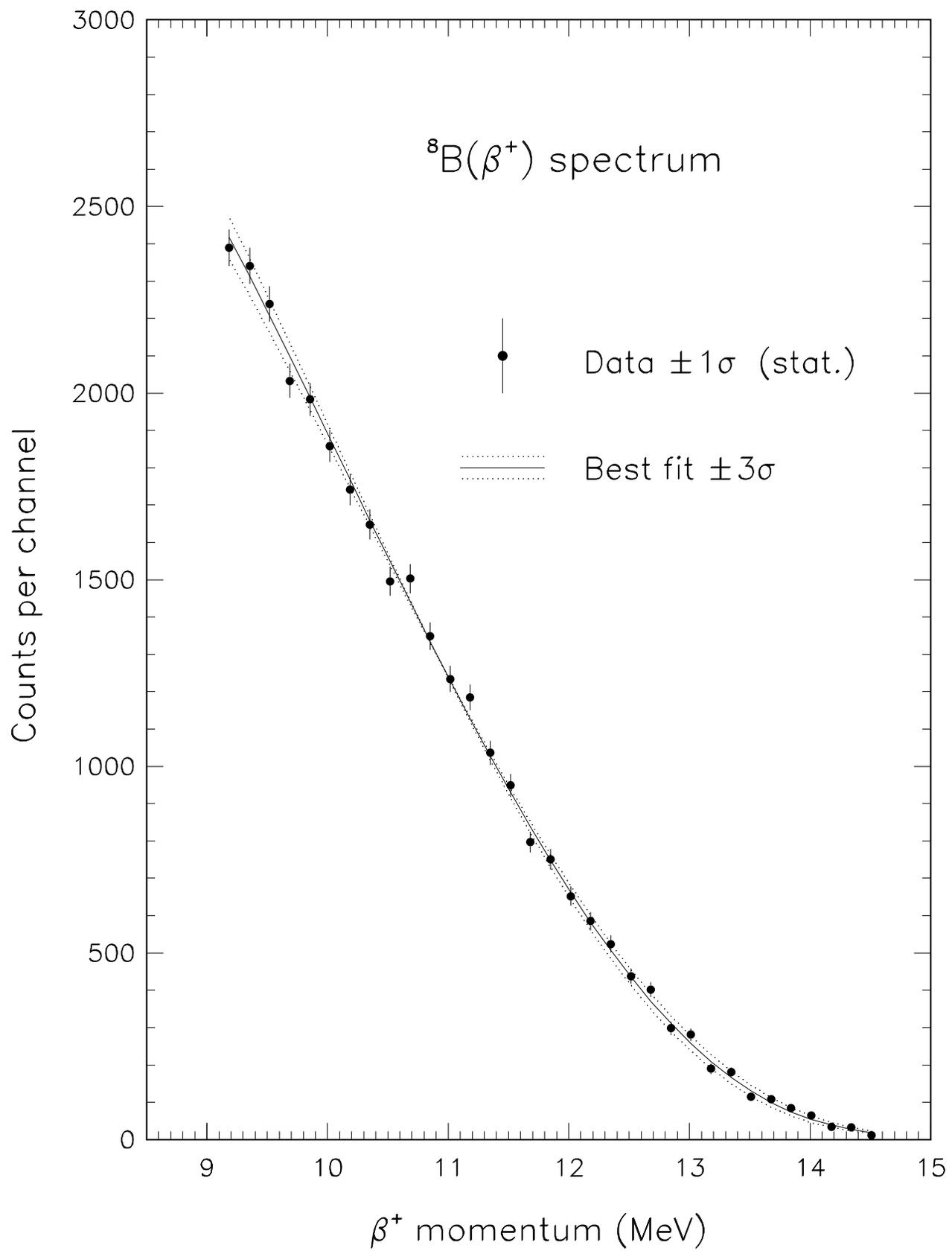
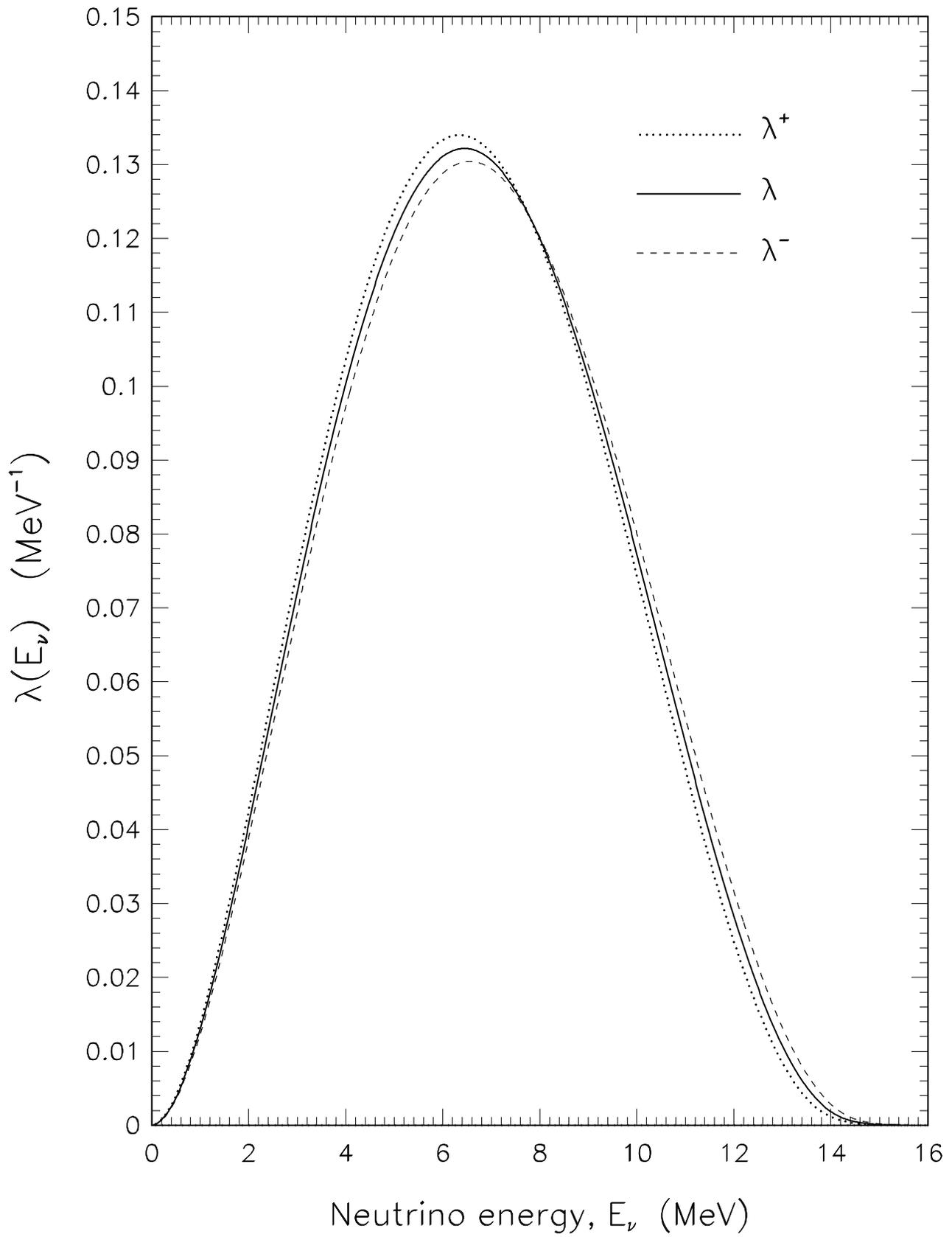
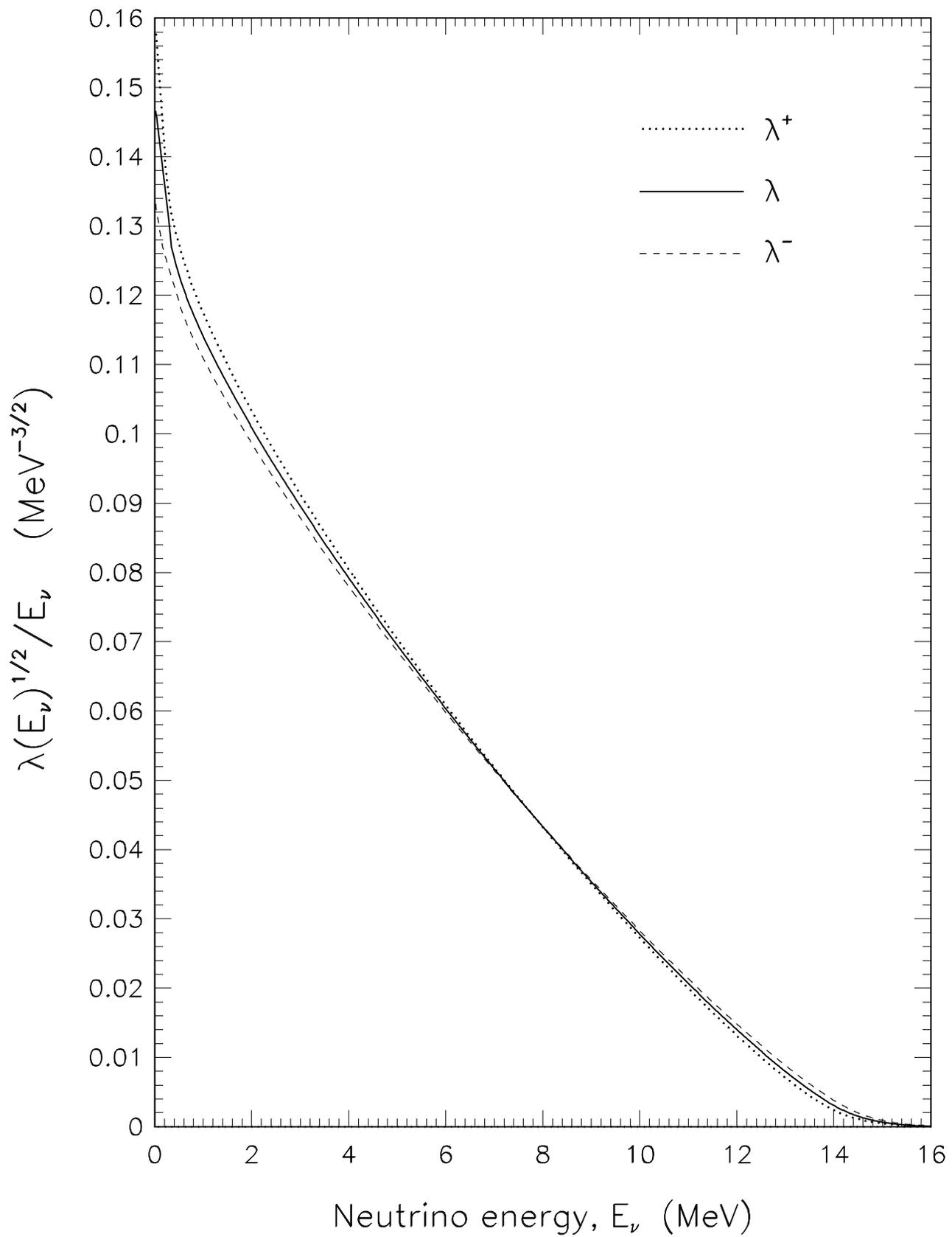
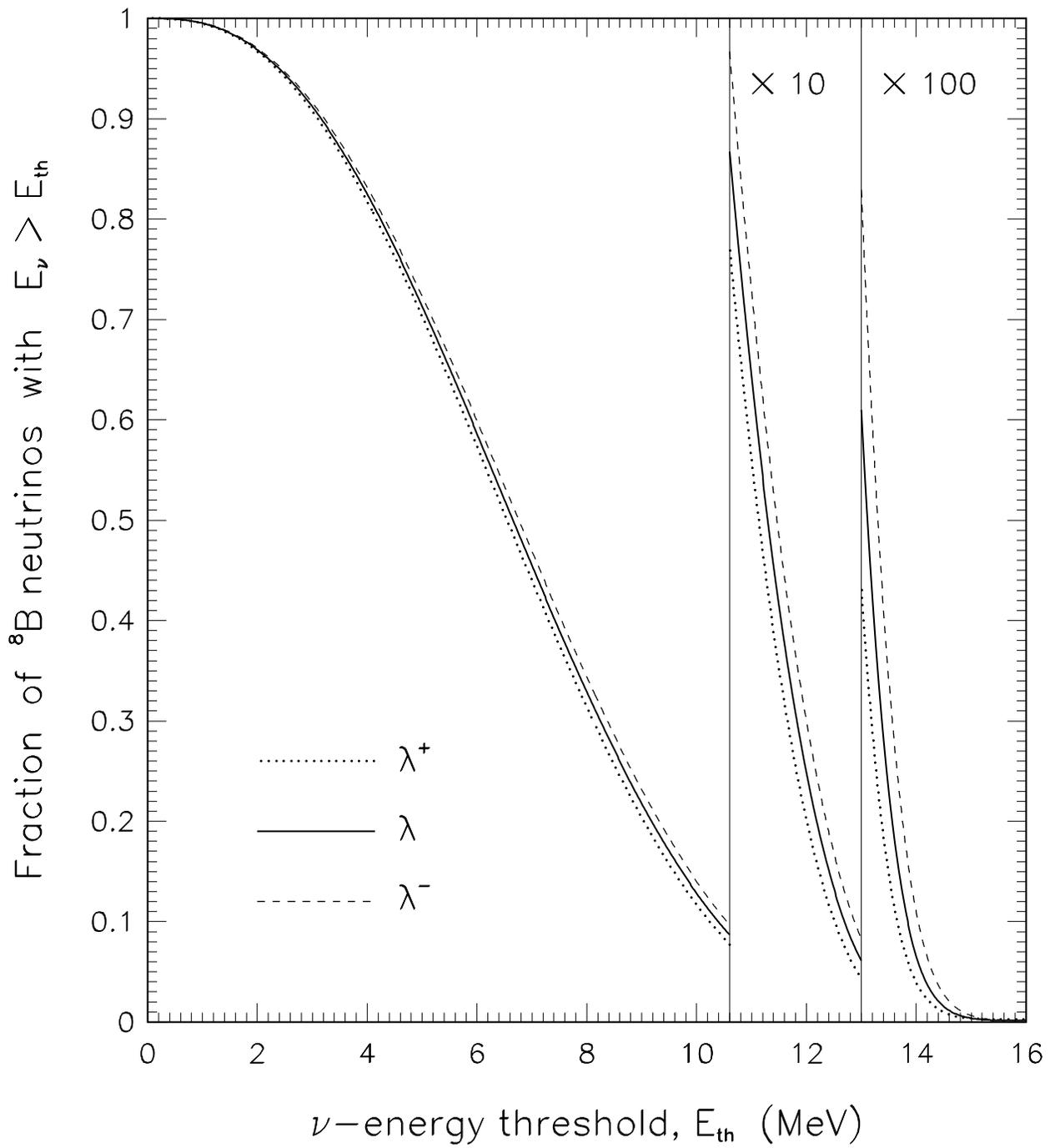
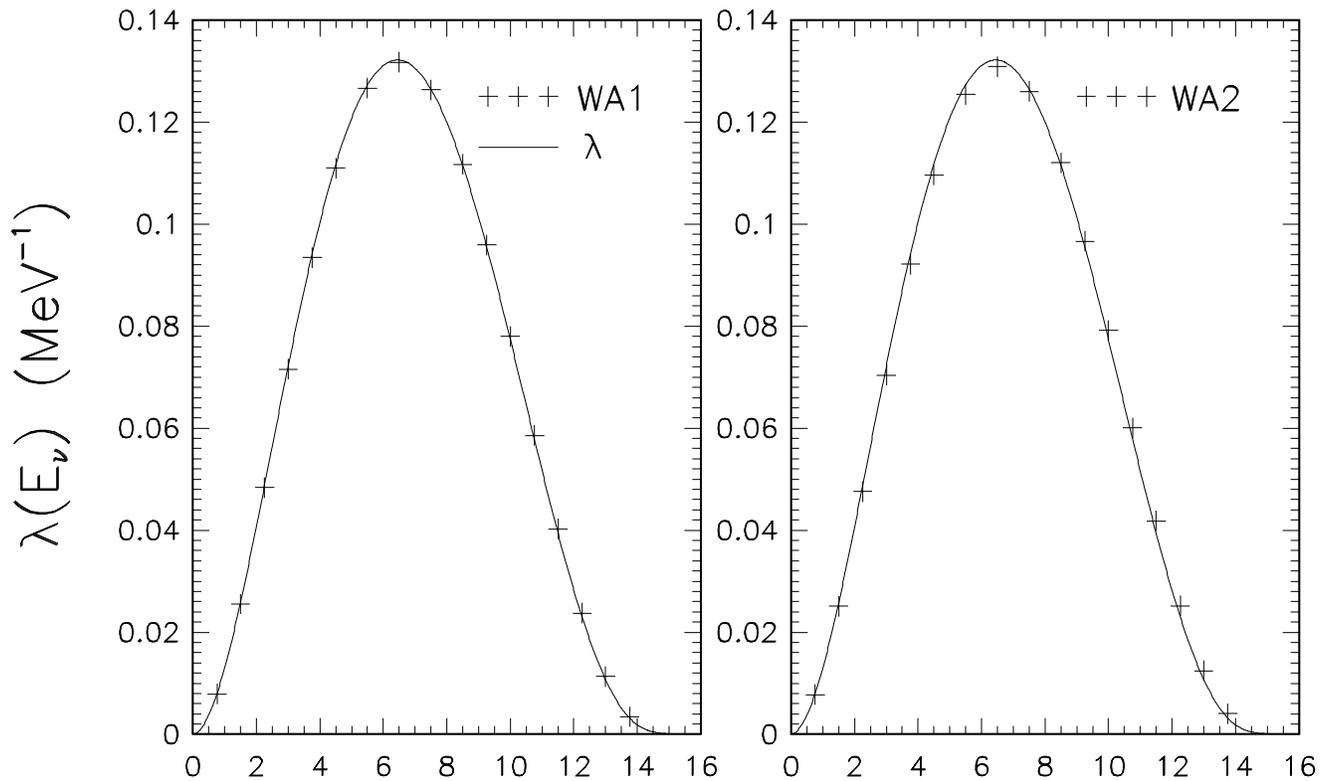
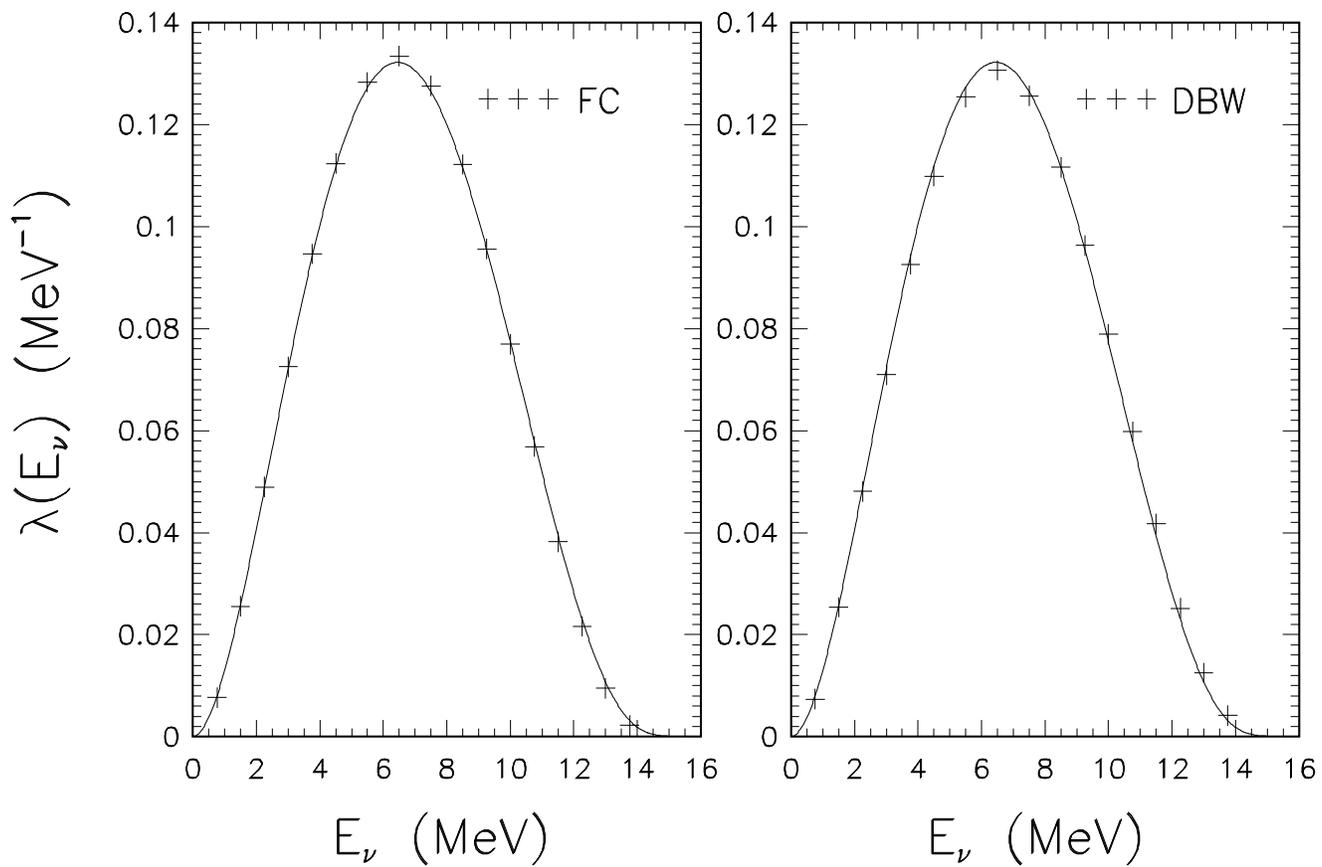
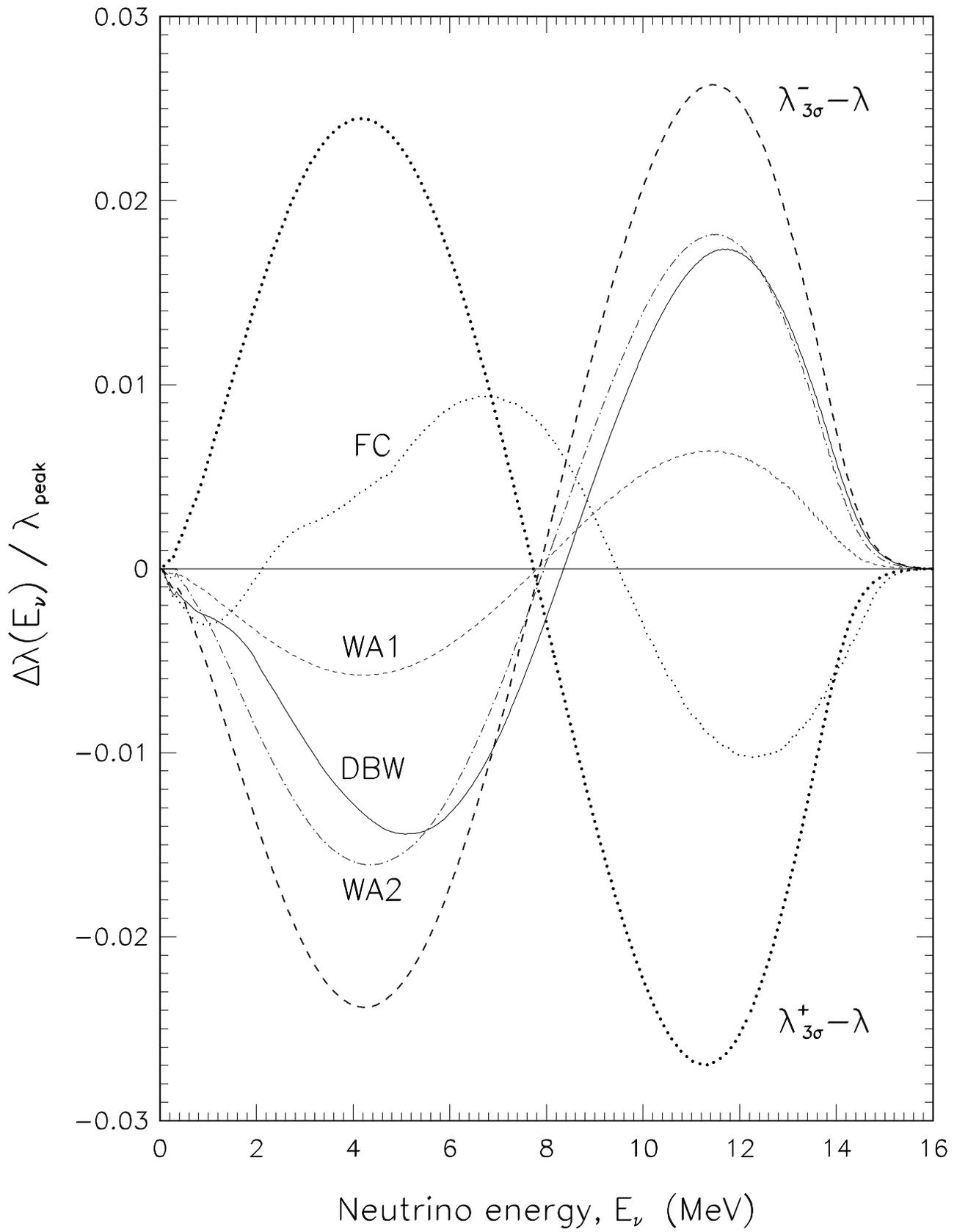
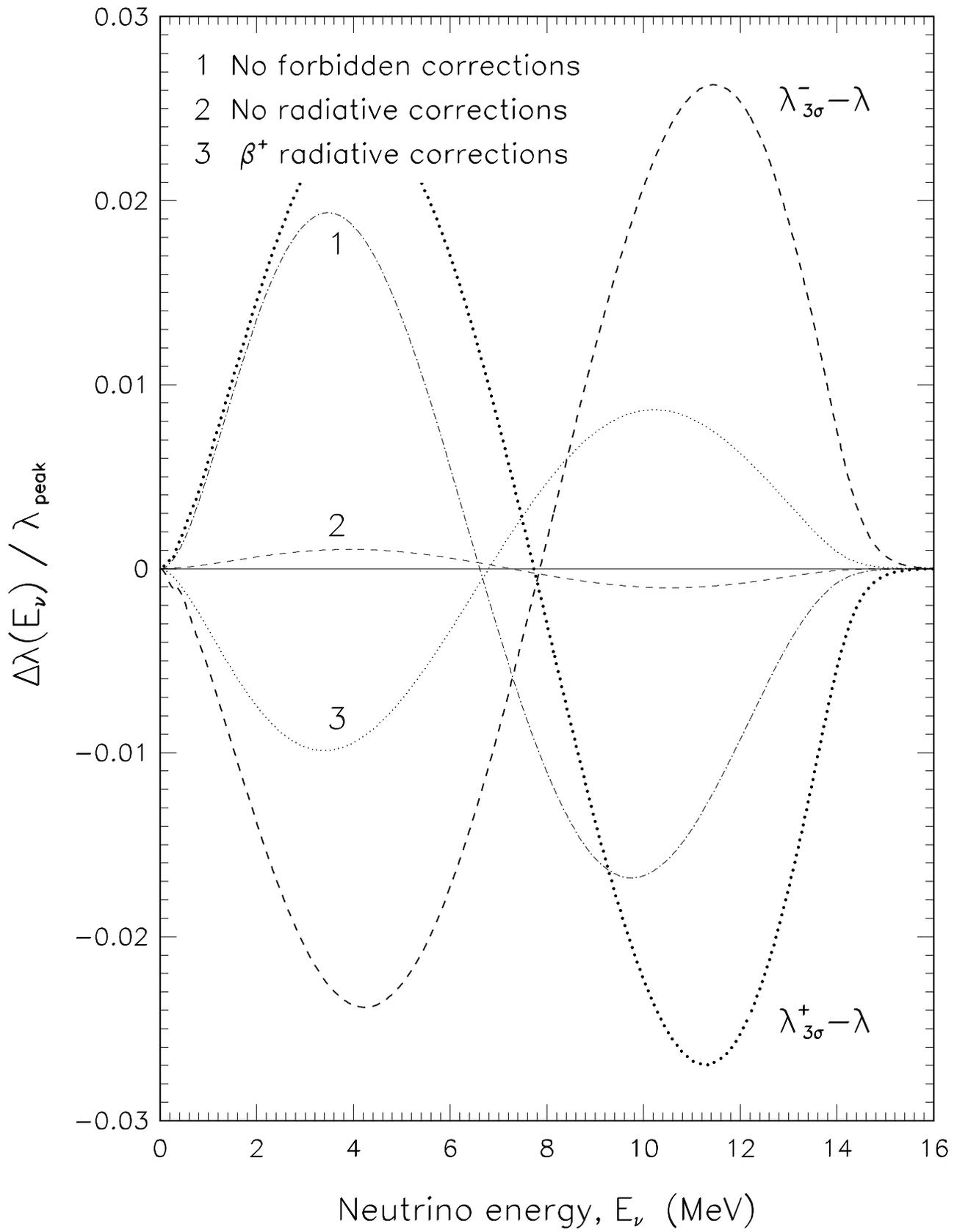
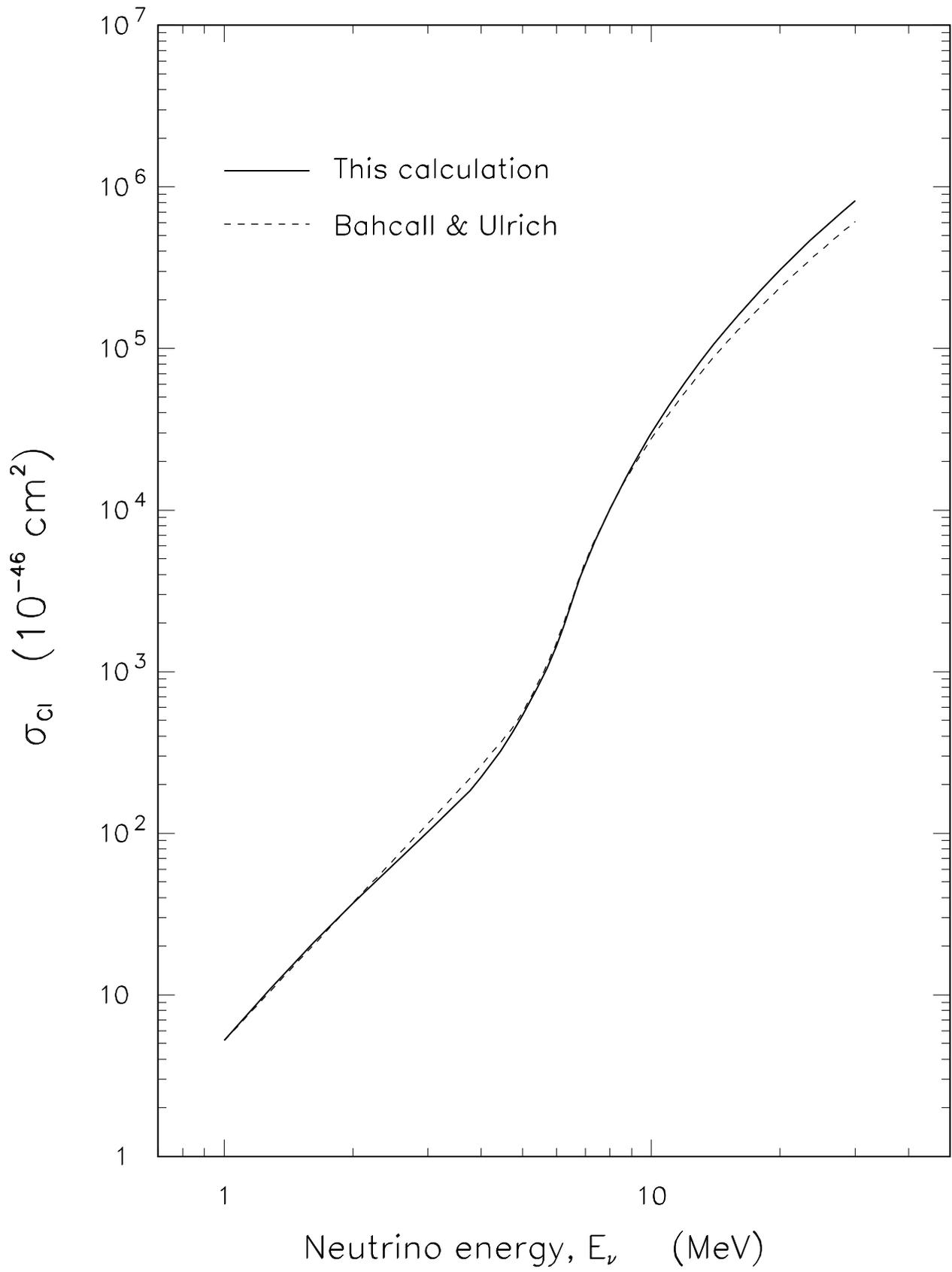
\end{document}